\documentclass[10pt,letterpaper,oneside]{article}

\usepackage{amssymb,amsmath,amsfonts,amsbsy,setspace}
\usepackage{graphicx,verbatim}

\doublespace

\usepackage{epsf,wrapfig,epsfig}

\newtheorem{defn}{Definition}[section]
\begin{document}

\title{Computation of Finite Time Lyapunov Exponents using the Perron-Frobenius operator}
\author{P. Tallapragada}
\date{}

\maketitle

\begin{abstract}
The problem of phase space transport which is of interest both theoretically and from the point of  view of applications has been investigated extensively using geometric and probabilistic methods. Two of the important tools for this that emerged in the last decade  are the Finite time Lyapunov exponents (FTLE) and the Perron-Frobenius operator. The relationship between these approaches  has not been clearly understood  so far. In this paper a methodology is presented to compute the FTLE from the Perron-Frobenius operator, thus providing a step towards combining both the methods into a common framework.
\end{abstract}
\textbf{Transport on finite time scales in dynamical systems has been studied intensively in the last two decades. A significant motiviation for this study has been transport in fluid systems varying from micro fluidics to geophysical flows. These studies received a significant boost with the development of the concepts of finite time Lyapunov exponents (FTLE)  and Lagrangian coherent structures (LCS), \cite{Haller3}, \cite{Shadden1}, \cite{Shadden2}, which act as transport barriers. A probablistic approach to transport in dynamical systems developed in the last decade, which uses the Perron-Frobenius (PF) operator, \cite{Lasota}, to identify the so called almost invariant sets, \cite{Dellnitz}, \cite{Froyland1}. The relationship between the two methods has not been well understood though both methods often yield roughly the same  results, \cite{Froyland2}. In this paper a method to compute the FTLE from the PF  operator is given by using an alternative  definition of the FTLE based  on the covariance of probability density functions. This method of computing the FTLE from the PF operator eliminates  the need for long time integrations of trajectories, often a source of errors. The method is illustrated with two simple examples  of  two dimensional fluid systems. The redefinition of the  FTLE along with its computation from the PF operator are a step towards combining the geometric and probabilistic methods into a common framework. }
\section{Introduction}
The problem of phase space transport has important applications such as in mixing and separation problems in fluid flows that vary in scale from the micro to the geophysical, interplanetary transport and instability of mechanical systems, to name a few. A variety of dynamical systems methods have been studied over the past three decades to explain transport mechanisms, to detect barriers to transport, and to quantify transport rates, see eg  \cite{Haller3}, \cite{Shadden1}, \cite{Shadden2},  \cite{Dellnitz}, \cite{Froyland1}, \cite{Wiggins_book2}, \cite{WigginsIde}, \cite{Wiggins1}, \cite{Froyland3}, \cite{Froyland4}, \cite{DuToit_thesis}.  These methods fall into two main categories, the geometric and the probabilistic. Under the umbrella  of geometric methods are the techniques of invariant manifolds (of fixed points) , lobe dynamics and Finite time Lyapunov exponents (FTLE) and Lagrangian coherent structures (LCS). The method of  FTLE and  LCS has proven to be particularly useful in studying transport in time  dependent systems and has found a variety of  applications, for e.g., \cite{Lekien2},  \cite{PhaniRoss}, \cite{Haller2}, \cite{Ross_Lekien}, \cite{Tang_Haller} and \cite{Peng_Dabiri2}. The probabilistic approach studies the transport of densities and the so called almost invariant sets and coherent sets. These methods too have been successfully applied in the study of various geophysical flow problems, \cite{Froyland4} and mixing in micro channels \cite{Grover_Paper}.   

The method of LCS studies stretching and contraction around reference trajectories. The  LCS method is therefore local in nature; it provides information about invariant manifolds that determine transport in phase  space. The current method of LCS relies on computing the FTLE field using long time computations, since LCS usually can be identified only after a significant time of integration. The  disadvantage of this is that, excessive stretching of line elements can introduce  computational errors. Moreover specific checks on whether the  stretching of line  elements is within the linearized regime can be difficult to incorporate in the current algorithms on computing the  FTLE. The probabilistic method on the other hand ignores the local transport structures, but using the transfer operator divides the  phase space into maximally invariant sets. There have  been a few attempts to explore the relationship between the geometric and  probabilistic descriptions of  phase space transport, such as \cite{Froyland2} and \cite{PhaniRoss2}. The aim of this short paper is to present a technique of computing the FTLE using the Perron-Frobenius operator that is a step towards combining the geometric and probabilistic methods, by making the Perron-Frobenius operator the common tool to both. By utilizing the Perron-Frobenius operator to compute the FTLE, this method also strengthens the probabilistic interpretation of the FTLE, identified in \cite{PhaniRoss2} and \cite{Phani_thesis}. This approach has  the added benefits of eliminating the issue of linearization around a reference trajectory in the existing formulation of the FTLE and reduces the time to compute the FTLE field for time-dependent and periodic flows.
\section{Review of FTLE and LCS}
The formulation of FTLE and the Perron-Frobenius operator is reviewed in this section. This review is intended to provide an intuitive background and set the context for the computation of the FTLE using the Perron-Frobenius operator. For the rigorous definitions and details on these methods,  the reader is referred to \cite{Dellnitz}, \cite{Froyland1}, \cite{Shadden1}, \cite{Shadden2}, \cite{Haller3}.
\subsection{Finite Time Lyapunov exponents}
Let $M \subset \mathbb{R}^n$ be compact and $\phi_{t_0}^t (x) : M \times \mathbb{R} \times \mathbb{R}  \mapsto M$ be a smooth flow on $M$. Let the associated vector field be $\dot{\mathbf{x}} = F(\mathbf{x},t)$ for $\mathbf{x} \in M$. Consider a reference trajectory passing through the point $\mathbf{x}(t_0)$  and a perturbed trajectory passing through the point $\mathbf{x}(t_0) + \delta \mathbf{x}(t_0)$ at time $t_0$. The flow $\phi_{t_0}^{t}$ maps these points to $\phi_{t_0}^{t}(\mathbf{x}(t_0))$ and $\phi_{t_0}^{t}(\mathbf{x}(t_0) + \delta \mathbf{x}(t_0))$ at time $t$ and the perturbation grows to $\delta \mathbf{x}(t)$.

Expanding $\phi_{t_0}^{t}(\mathbf{x}(t_0) + \delta \mathbf{x}(t_0))$  in a Taylor series about the point $\mathbf{x}(t_0)$ we get 
\begin{equation}\label{eq:taylor_perturb}
\delta \mathbf{x}( t) = \phi_{t_0}^{t}(\mathbf{x}(t_0)) -\phi_{t_0}^{t}(\mathbf{x}(t_0) + \delta \mathbf{x}(t_0)) = \frac{d \phi_{t_0}^t}{d \mathbf{x}} \delta \mathbf{x}(t_0) + O(\left\|\delta\mathbf{x}(t_0)\right\|^2) 
\end{equation} 
The norm or magnitude of $ \delta \mathbf{x}(t_0 + t) $ can be found using the standard inner product on $\mathbb{R}^n$.
\begin{equation}
\left\| \delta \mathbf{x}(t) \right\| = \sqrt{ \left\langle  \frac{d \phi_{t_0}^t}{d \mathbf{x}}\delta \mathbf{x}(t_0), \frac{d \phi_{t_0}^t}{d \mathbf{x}}\delta \mathbf{x}(t_0)   \right\rangle}   = \sqrt{ \left\langle \delta \mathbf{x}(t_0), \left(\frac{d \phi_{t_0}^t}{d \mathbf{x}}\right)^* \frac{d \phi_{t_0}^t}{d \mathbf{x}}\delta \mathbf{x}(t_0)   \right\rangle }
\end{equation}
where $*$ denotes the transpose and the gradient $\left(\frac{d \phi_{t_0}^t}{d \mathbf{x}}\right)$ is evaluated at $\mathbf{x}(t_0)$. 
The maximum growth of a perturbation is therefore given by the maximum principal stretch, i.e., by the maximum eigenvalue of $\mathbf{C}$.
\begin{equation}\label{eq:max_growth}
max \left\| \delta \mathbf{x}(t) \right\| = \sqrt{\lambda_{max}(\mathbf{C}(\mathbf{x}(t_0), t_0,t)} \left\| \delta \mathbf{x}(t_0) \right\| 
\end{equation}
where $\mathbf{C}(\mathbf{x}(t_0), t_0,t) =  \left(\frac{d \phi_{t_0}^t}{d \mathbf{x}}\right)^* \left(\frac{d \phi_{t_0}^t}{d \mathbf{x}} \right)$ is the Cauchy-Green tensor. The growth in the perturbation depends on the initial point $\mathbf{x}$, initial time $t_0$ and the evolution or integration time $T = t-t_0$.
\begin{defn}\label{defn:FTLE_defn1}
The maximum FTLE is defined as, \cite{Shadden1}, \cite{Shadden2},
\begin{equation}\label{FTLE_defn_eqn}
\sigma(\mathbf{x}(t_0),t_0,T) = \frac{1}{T} \log \left( \sqrt{\lambda_{max}(\mathbf{C}(\mathbf{x}(t_0),t_0,t))} \right)
\end{equation}
\end{defn}
The leading FTLE gives the time averaged rate of linearized stretching in a neighborhood around a reference trajectory. It is intuitively clear that regions of the phase space with locally high values of FTLE will stretch and separate. The sets with high FTLE act as repelling barriers in the flow. This intuitive idea of barriers is formalized by the concept of Lagrangian coherent structures (LCS), defined as ridges in scalar FTLE field, \cite{Shadden1}, \cite{Shadden2}. Ridges can be defined precisely by appealing to differential geometric quantities as in \cite{Eberly}, \cite{Shadden1} and \cite{Shadden2}.

\subsection{Perron-Frobenius operator} \label{sec:PF_op}
Let $\mu$ denote the Lebesgue measure and let $B$ be a measurable set and $f \in L^1$ be a probability density function, $L^1$ being the space of  Lebesgue measurable functions. The unique operator $\mathcal{P}_{t_0}^t:L^1 \mapsto L^1$ defined  by
\begin{equation} \label{eq:Perron-Frobenius_op}
\int_{B}\mathcal{P}_{t_0}^tf d\mu =  \int_{(\phi^{t}_{t_0})^{-1}(B)} f d\mu
\end{equation} 
 is called the Perron-Frobenius operator for the flow $\phi_{t_0}^t$, \cite{Lasota}. Equation \ref{eq:Perron-Frobenius_op} which holds for all $\mu$ measurable sets follows from the Radon-Nykodym theorem.

In practice it is usually necessary to numerically approximate the operator $\mathcal{P}_{t_0}^t$. This is done by discretizing the domain, $M$, into a finite number of sets, say $\left\{ B_1, B_2, ..., B_n\right\}$ which is essentially a grid of boxes. A projection $\pi : L^1 \mapsto span\left\{B_1, . . ., B_n\right\}$ defined by $\pi f = \sum_{i=1}^{n} c_i \mathcal{X}_{B_i}$, where $\mathcal{X}_{B_i}$ is the characteristic function of the set $B_i$ and $c_i = \frac{\int f d\mu}{\mu(B_i)}$ gives a finite dimensional approximation of $f$. Since $f$ is a probability density function $c_i = \frac{1}{\mu(B_i)}$. Similarly $\mathcal{P}_{t_0}^tf$ is projected on $span\left\{B_1, . . ., B_n\right\}$. The operator $P_{t_0}^t : \pi f \mapsto \pi \mathcal{P}_{t_0}^t f$ is a linear operator between finite dimensional vector spaces. Further taking the box measures $\mu (B_i) = \mu (B_j)$ for all $i, j \in [1,n]$, $P_{t_0}^t$ becomes a stochastic transition matrix. The entries of the matrix $P$ are determined by a Monte-Carlo simulation \cite{Dellnitz} and \cite{Froyland1}. Each box in the domain contains a fixed number of points (initial conditions), which are integrated from a time $t_0$ to $t$. The final position of the points gives the matrix $P$ as -
\begin{equation}\label{eq:P_mat}
(P_{t_0}^t)_{ij} = \frac{\mu(B_i \cap (\phi^t_{t_0})^{-1}(B_j))}{\mu(B_j)}
\end{equation}
A time reversible operator $P$ is required to apply the above definition for flows in forward time, \cite{Froyland1}. This is achieved by creating a reversible Markov operator $P_r$ given by
\begin{equation} \label{eq:P_reverse}
(P_{t_0}^t)_r = \frac{(P_{t_0}^t) + \overline{(P_{t_0}^t)}}{2}
\end{equation}
where $\overline{P}$ is the time reversed analogue of $P$. Its elements are given by
\begin{equation}\label{eq:P_bar}
(\overline{P_{t_0}^t})_{ij} = \frac{u_{1j}(P_{t_0}^t)_{ji}}{u_{1i}}
\end{equation}
where $u_{1j}$ and $u_{1i}$ are components of the first left eigenvector, $u_1$, of $P^t_{t_0}$. For a volume preserving flow in which the domain is uniformly discretized, $\overline{P_{t_0}^t} = (P_{t_0}^t)^{*}$, the transpose of $P_{t_0}^t$. Henceforth for convenience the time reversible operator $(P^t_{t_0})_r$ is referred to simply as $P$ unless  there is an ambiguity about the initial and final times $t_0$ and $t$.  The Markov operator $P$ has the semigroup property of $P_{t_0}^t =  P_{t_0}^s  P_{s}^t$, where $s \in (t_0,t)$. This property is  used to simplify the computations of the  FTLE.

\section{Computation of FTLE using the Perron-Frobenius operator}
\subsection{Definition of set oriented FTLE}
\begin{figure}
\begin{center}
\begin{tabular}{cc}
\includegraphics[width = 0.47\textwidth]{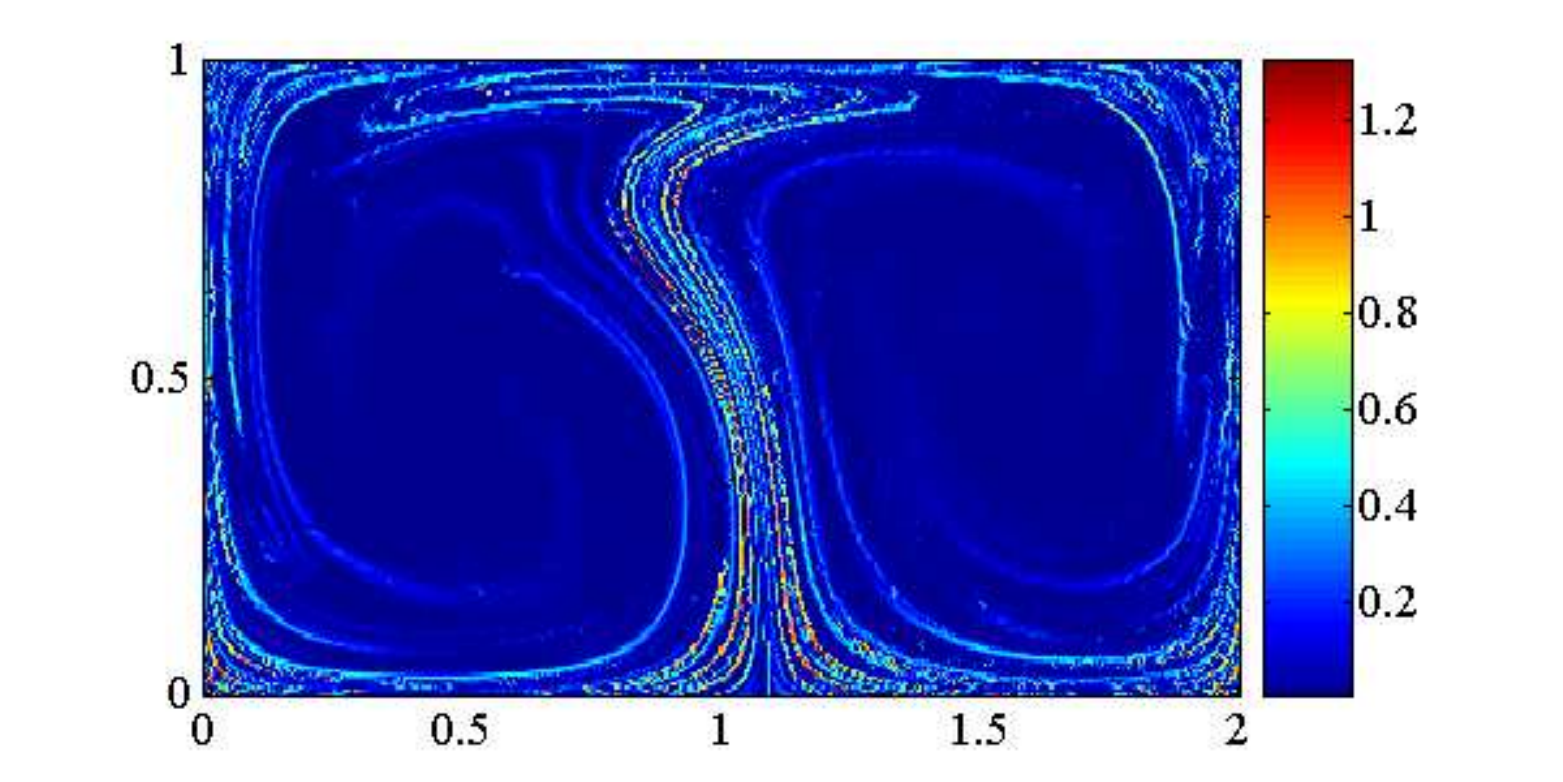} &
\includegraphics[width = 0.47\textwidth]{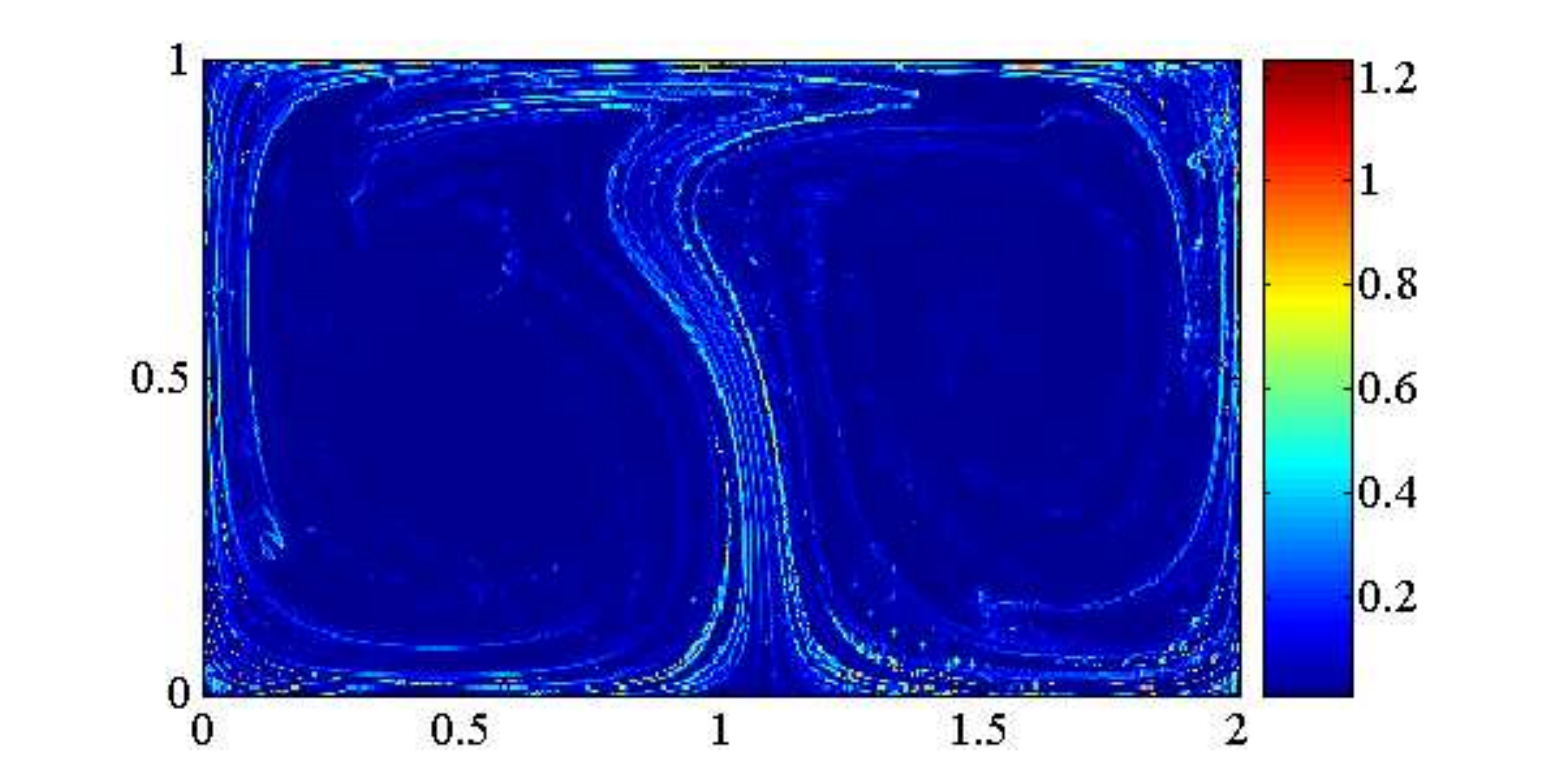} \\
\footnotesize{(a) Magnitude of first-order terms in  eq. (\ref{eq:taylor_perturb2}) }&
\footnotesize{(b) Magnitude of second-order terms in  eq. (\ref{eq:taylor_perturb2})}
\end{tabular}
\end{center}
\caption{\label{fig:fig_first_second_order}{\footnotesize{(a) Comparison of first- and second-order terms in eq. (\ref{eq:taylor_perturb2}) for $T = t-t_0= 10$.}}}
\end{figure}

The concepts of FTLE and LCS reviewed previously have been used fruitfully in many areas as has been pointed out earlier. However the standard computational implementation of FTLE, using  finite differences, suffers from some drawbacks, stemming from the integration time $T$ and the linearization around reference trajectories. In equation \ref{eq:taylor_perturb}, it is assumed  that the second (and higher ) order terms are negligible. However the magnitude of  the higher  order terms depends on the evolution time $T = t - t_0$. If the evolution time $T$ is too high then the higher order terms may be  comparable to the first order terms in equation \ref{eq:taylor_perturb}. If the evolution time $T$ is too low, then one cannot detect any interesting structure in the FTLE field. Often, the evolution time $T$ is selected in a subjective  fashion without checking the validity of  equation \ref{eq:taylor_perturb}. The problem with this is illustrated with the  example  of the double gyre flow, a prototype in the LCS literature, \cite{Shadden1, Froyland2}.  The double gyre flow is defined by the stream function $\psi(x,y,t) = A \sin(\pi f(x,t))\sin{(\pi y)}$, with  $f(x,t) = \epsilon x^2 \sin{\omega t} + x(1 - 2\epsilon\sin{\omega t})$, where we use the parameters $A=0.25$, $\omega = 2\pi$ and $\epsilon = 0.25$.  The time period of the flow is $\tau = 1$. Now consider the equation \ref{eq:taylor_perturb} with the second  order terms explicitly written,
\begin{equation}\label{eq:taylor_perturb2}
\delta \mathbf{x}(t_0 + t) = \phi_{t_0}^{t}(\mathbf{x}) -\phi_{t_0}^{t}(\mathbf{x} + \delta \mathbf{x}) = \frac{d \phi_{t_0}^t}{d \mathbf{x}} \delta \mathbf{x}(t_0) + \frac{1}{2}\delta \mathbf{x}(t_0)^* \frac{d^2 \phi_{t_0}^t}{d \mathbf{x}^2} \delta \mathbf{x}(t_0) + O(\left\|\delta\mathbf{x}(t_0)\right\|^3).
\end{equation}
In this equation it can be shown that the second and first order  terms are comparable in magnitude. The max-norm for matrices is used for this  comparison. Figure \ref{fig:fig_first_second_order} shows the plot of $\lambda_{\max}\left(\frac{d\phi^t_{t_0}}{d\mathbf{x}}\right) \| \delta\mathbf{x} \|$ and $\frac{1}{2}\lambda_{\max}\left(\frac{d^2\phi^t_{t_0}}{d\mathbf{x}^2}\right) \|\delta\mathbf{x}\|^2$. The magnitude of the second order terms is more than half that of the first order  terms in equation \ref{eq:taylor_perturb2}.  This is  true for a smaller evolution time as  well. The ridges in FTLE  field for the double gyre flow are generated by moving instantaneous stagnations points (ISPs) . Material line elements close to the ISPs stretch in a  shorter time as compared to material line elements farther away. Therefore ineresting structures such  as ridges in the FTLE field require a long period  of integration during which the nonlinear deformation of line elements close to the ISPs occurs. 

\begin{figure}[!h]
\begin{center}
\includegraphics[height = 0.25 \textwidth, width = 0.6\textwidth]{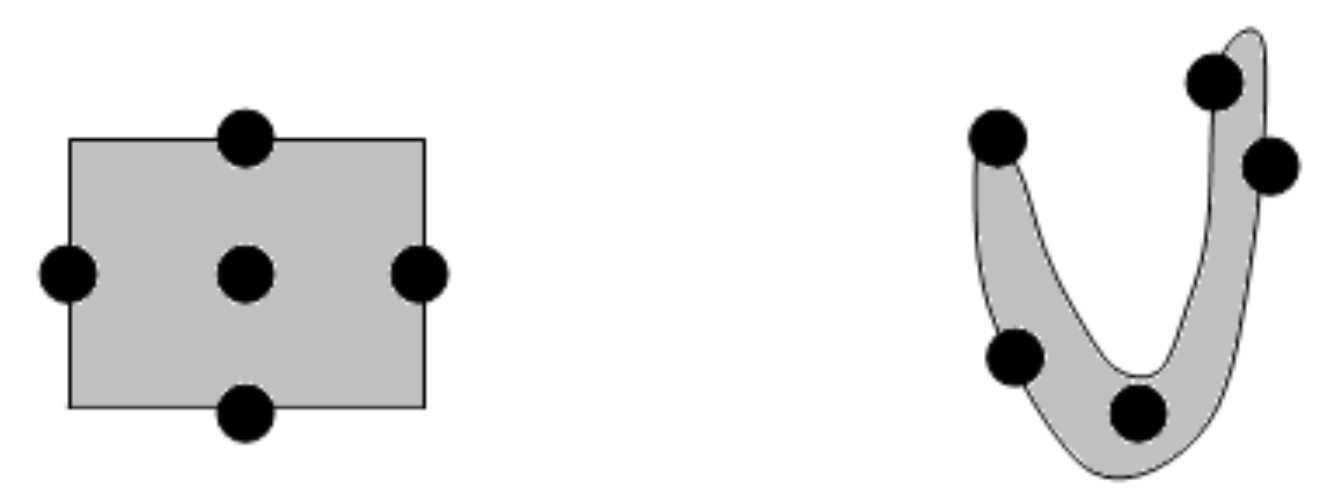}
\caption{\label{fig:fold_stretch}{\footnotesize{Folding of material elements.}}}
\end{center}
\end{figure}

To study the linearized flow in equation \ref{eq:taylor_perturb2}, numerics have to be carefully planned to ensure that the approximation of linearization is valid. Usually this is accomplished by selecting a very small $\delta \mathbf{x}$ and either by keeping the time of integration $T$ small enough or rescaling the perturbation as it grows very large. If perturbations around a specific trajectory grow too large and need to rescaled, then additional reference points have to be introduced in that region to obtain the FTLE field at a fine enough resolution. Alternatively the computations could start with a crude mesh of initial points and  refined  iteratively by introducing new initial points based on the finite time stretching, requiring adaptive meshing of initial conditions, which was explored in \cite{Ross_Lekien}.  However such techniques  of mesh refinement are  not based on the magnitude of the second order terms in  equation \ref{eq:taylor_perturb2}. Beyond the  fact that numerical  algorithms for the calculation of FTLE do not check the  validity of the  linearization of the flow,  there is the  additional fact that interesting structures in the FTLE field develop for longer eveolution times during which material lines stretch nonlinearly and even fold as shown in figure \ref{fig:fold_stretch} which was discussed earlier in \cite{Ross_Lekien}. 

The difficulties in the computation of the FTLE can  be overcome with a set oriented definition of the FTLE. Instead of tracking only a reference trajectory and the adjacent nodes of the finite difference mesh, one tracks the movement of the whole set (shown in gray) in figure \ref{fig:fold_stretch} and  consider the distribution of points in the set, then folding and nonlinear deformation can be accounted for. A new definition of the FTLE was proposed in \cite{PhaniRoss2} which computed the deformation of sets instead of the stretching of line elements and partially addressed  the concerns of second order terms, integration time and folding of material lines. We review the new definition here.

\begin{figure} [!h]
\begin{center}
\includegraphics[height = 0.23 \textwidth, width = 0.75\textwidth]{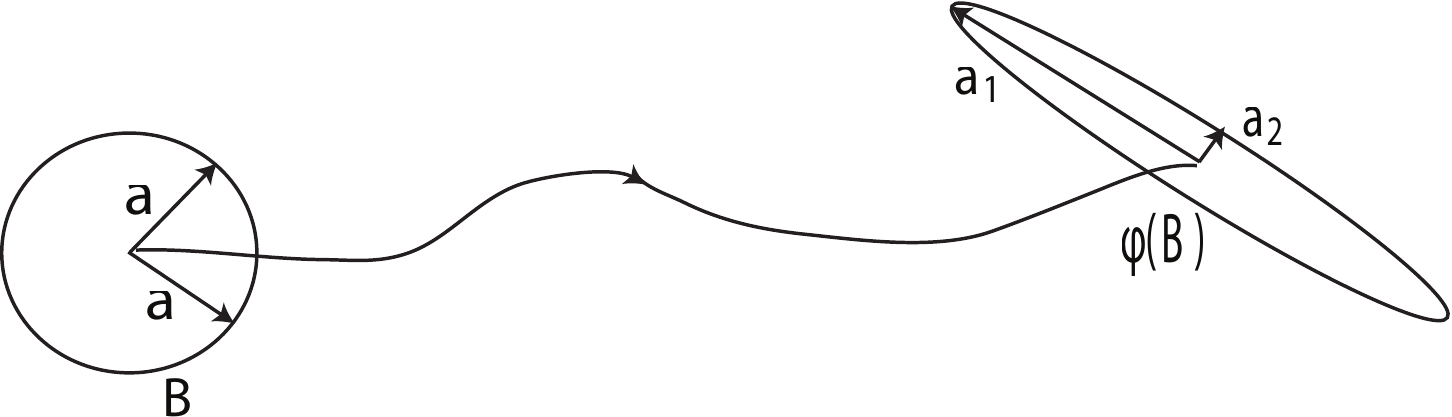}
\caption{\label{fig:fig_blob_deform}{\footnotesize{Deformation of a blob under the flow}}}
\end{center}
\end{figure}

For illustrating the concept we assume the flow  $\phi_{t_0}^t  : M  \mapsto M$ is is  two dimensional with  $M \subset \mathbb{R}^2$. The method of computing FTLE using the SVD of the Cauchy-Green tensor essentially computes the linearized stretching of a neighborhood, a set $B$, under the action of the flow $\phi_{t_0}^t$ as shown in figure \ref{fig:fig_blob_deform}. The FTLE for the reference trajectory in this case is $\sigma = \frac{1}{T} \log{\left(\frac{a_1}{a}\right)}$ where $T$ is the time of evolution of the trajectory. 

In the standard FTLE method the evolution of  the set  $B$ is assumed to be determined by the evolution of two vectors, the directions of principal stretches. For the set oriented  definition the deformation of the set $B$ is tracked by the evolution of a random vector $X=  [X_1, X_2]$ defined by a probability density function $f(x_1, x_2)$ which is initially a uniform probability density function supported on $B$ given by, $f = \frac{1}{\mu(B)}\mathcal{X}_B$,  where $\mathcal{X}_B$ is characteristic function of $B$. The covariance matrix of $f$ is $I_{ij} = E[(X_i - \overline{X}_{i})(X_j - \overline{X}_{j})]$, with $i = 1, 2$ and $j = 1, 2$ where $[\overline{X}_{1}, \overline{X}_{2}]$ is the mean value of the  random vector $X$ and $E[\cdot]$ denotes the expectation with $E[X] = [\int x_1 f(x_1, x_2) d\mu, \int x_2 f(x_1, x_2) d\mu]$. Under the action of the flow $\phi_{t_0}^t$, $f$ is mapped to $\mathcal{P}_{t_0}^t f$ where $\mathcal{P}_{t_0}^t$ is the associated Perron-Frobenius operator. In matrix notation the covariance  $I_{ij}$ is 

\begin{eqnarray}
I =  
\left(\begin{array}{cc}
	E[(X_1 - \overline{X}_{1})(X_1 - \overline{X}_{1})] & E[(X_1 - \overline{X}_{1})(X_2 - \overline{X}_{2})]\\
	E[(X_2 - \overline{X}_{2})(X_1 - \overline{X}_{1})] & E[(X_2 - \overline{X}_{2})(X_2 - \overline{X}_{2})]
\end{array}\right)
\end{eqnarray}

\begin{defn}\label{defn:FTLE_new_defn}
Let $I(f)$ be the covariance of $f$ and $I(Pf)$ the covariance of $Pf$ and let $\lambda_{max}(I)$ denote the maximum eigenvalue of $I$. Then the FTLE of $B$ denoted by $\sigma_I(B,t_0,t)$ is defined as -
\begin{equation}\label{eq:new_FTLE}
\sigma_I(B, t_0, t) = \frac{1}{t-t_0}\log{\left(\frac{\sqrt{\lambda_{max}(I(Pf))}}{\sqrt{\lambda_{max}(I(f))}}\right)}
\end{equation}
\end{defn}
It can be shown that by direct calculation that the covariance FTLE obtained from Definition 3.1 and the standard FTLE have the same value if the second and higher order terms in equation \eqref{eq:taylor_perturb2} are negligible when compared to the first order terms. In this case an initial circular blob deforms into an ellipse as shown in figure \ref{fig:fig_blob_deform}. The eigenvalues of the $\mathbf{C}$ are $a_1^2$ and $a_2^2$. The standard FTLE is $\sigma = \frac{1}{t-t_0} \log{\frac{a_1}{a}}$. The covariance matrix for the deformed ellipse is 
\begin{eqnarray}
I(Pf) = 
\left(\begin{array}{cc}
	 \frac{1}{4}\pi a_1^3a_2 &  0\\
	0 & \frac{1}{4}\pi a_1a_2^3
\end{array}\right)
\end{eqnarray}
giving $\lambda_{max}(I(Pf)) = \frac{1}{4}\pi a_1^3a_2$. The covariance matrix $I(f)$ is a diagonal matrix with $I(f)_{11} = I(f)_{22} = \frac{1}{4}\pi a^4$ giving $\lambda_{max}(I(f)) = \frac{1}{4}\pi a^4$. This gives the covariance FTLE, $\sigma_I = \frac{1}{t-t_0} \log{\left(\frac{\sqrt{a_1^3a_2}}{\sqrt{a^4}}\right)}$. For volume  preserving flows,  $\pi a_1a_2 = \pi a^2$. This gives the covariance FTLE, $\sigma = \frac{1}{t-t_0} \log{\frac{a_1}{a}}$, the same value as  the standard FTLE.
The standard FTLE is obtained by the stretching of  the value of the covariance based FTLE is the same as that obtained from traditional FTLE  calculation using line stretching.

Since $I$ is the covariance of $f$, it provides a probabilistic interpretation of the FTLE. Simultaneously $I$ can also be interpreted as the moment of inertia of the set $B$ and provides a geometric description of the deformation or distortion of the set. The definition of $\sigma_I$ avoids the linearization of the flow and the computation of the stretching of line elements. Further it is a set-oriented method and directly computes the deformation of a set instead of inferring it from the deformation of line elements. If the linearized flow is valid the covariance FTLE and the standard FTLE are equal. However the covariance FTLE is a better measure  of finite deformation of sets. The covariance based method of FTLE is a bridge between the geometric approach of measuring line stretching and the probabilistic approach of almost invariant sets.

In practice any set $B$ and a density function $f$ supported on $B$ has to be approximated by discrete points. If the function $f = \frac{1}{\mu(B)}\mathcal{X}_B$ is a uniform density function, then $B$ can be approximated  by $N$ points, with coordinates $(x_{1i}, x_{2i})$, $i$ ranging from 1 to $N$. Each of these $N$ points has discrete measure  $\frac{1}{N\mu(B)}$. Similarly $\phi_{t_0}^t(B)$ is also approximated by the images of the $N$ points, $(\phi_{t_0}^t(x_{1i}, x_{2i}))$, each with measure $\frac{1}{N\mu(\phi_{t_0}^t(B))}$. For volume preserving flows this is the same as $\frac{1}{N\mu(B)}$. The covariance for the function $Pf$, $I(Pf)$, is approximated by

\begin{eqnarray} \label{eq:covar_Pf}
I(Pf) =  \frac{1}{N\mu(\phi_{t_0}^t(B))}
\left(\begin{array}{cc}
	\sum^{N}_{i=1} (\phi_{t_0}^t(x_{1i}) - \overline{\xi}_1))^2& \sum^{N}_{i=1} (\phi_{t_0}^t(x_{1i}) - \overline{x}_1)(\phi_{t_0}^t(x_{2i}) - \overline{\xi}_2)\\
	\sum^{N}_{i=1} (\phi_{t_0}^t(x_{1i}) - \overline{\xi}_1)(\phi_{t_0}^t(x_{2i}) - \overline{\xi}_2) & \sum^{N}_{i=1} (\phi_{t_0}^t(x_{2i}) - \overline{\xi}_2)^2
\end{array}\right)
\end{eqnarray}

where $[\overline{\xi}_1, \overline{\xi}_2]$ is the mean or expected value of the discrete approximation of $\phi_{t_0}^t(B)$. The initial covariance matrix can also be obtained similarly. The maximum eigenvalues of $I(f)$ and $I(Pf)$ can be plugged into equation \eqref{defn:FTLE_new_defn} to give the covariance FTLE.
\subsection{Computation of covariance FTLE using Perron-Frobenius operator}
The covariance FTLE is a set oriented redefinition of the FTLE. Therefore is natural to expect the Perron-Frobenius operator play a role in the computation of the covariance FTLE. This will be particularly useful since only short time integrations are necessary to calculate the Perron-Frobenius operator $P$. The operator $P_{t_0}^t$ can be  found  using a suitable set of short time intervals, $[t_0,t_1],[t_1,t_2], ..., [t_n,t]$; $P_{t_0}^t  =  P_{t_0}^{t_1}  P_{t_1}^{t_2}  ...  P_{t_n}^{t}$. Let $f$ be a uniform density function supported on a set $B$. Then as before $P_{t_0}^t$ maps $f$ under the action of the  flow to $P_{t_0}^tf$, $P_{t_0}^t: f \mapsto P_{t_0}^t f$. For time independent flows or periodic flows with period $\Delta t$, this becomes particularly easy, with $P_{t_0}^t  =  (P_{t_0}^{t_0 + \Delta t})^{n+1}$ where $\Delta t = t_{i+1} - t_i $ for $i = 0, 1, ..., n$. This avoids long time integration of trajectories. 

For two dimensional flows, $\phi_{t_0}^t  : \mathbb{R}^2 \mapsto \mathbb{R}^2$, the operator $\mathcal{P}_{t_0}^t$ is numerically approximated by  $P_{t_0}^t$ using the box discretization method described in section \ref{sec:PF_op}. Each box contains a fixed number of uniformly distributed points. The entries of the matrix $P_{t_0}^{t_0 + \Delta t}$ are calculated or a suitably chosen  $\Delta t$ and the time reversible operator is found from equations \eqref{eq:P_mat}, \eqref{eq:P_reverse} and \eqref{eq:P_bar}. To compute the FTLE over the entire domain, we take the  projection  of a uniform density function such that $\pi f = u^i=[0 \;0\; . . .\, 1/\mu(B_i)\; 0\; 0]$, with the $ith$ column being equal to the inverse of the measure of $B_i$.  The vector $u^i$ evolves to $v^i(t_0 + \Delta t)$ at  $t_0 + \Delta t$, given by,
\begin{equation}
v^i(t_0 + \Delta t) = u^i\;P_{t_0}^{t_0 + \Delta t} = \frac{1}{\mu(B_i)}(P_{t_0}^{t_0 + \Delta t})_{ij}
\end{equation}
Similarly the evolution of $u_i$ at time $t = t_0 + (n+1)\Delta t$ is given by
\begin{equation}\label{eq:v_t}
v^i(t) = u^i\;(P_{t_0}^{t_0 + \Delta t})^{n+1} = \frac{1}{\mu(B_i)}(P_{t_0}^{t_0 + \Delta t})_{ij}^{n+1}
\end{equation}
Equation \ref{eq:v_t} defines how a set of points uniformly distributed over  the set $B_i$ at $t_0$ is disperese at time $t$. The the covariance of this set can be found from the $ith$ row of $(P_{t_0}^{t_0 + \Delta t})^{n+1}$ as follows

\begin{eqnarray} \label{eq:covar_P_discrete}
I(v^i(t)) =  \frac{1}{\mu(B_i)}
\left(\begin{array}{cc}
	\sum^{N}_{j=1} (( \overline{\zeta}_1^j - \overline{\xi}_1^i)^2 (P_{t_0}^t)_{ij}) & \sum^{N}_{j=1} (( \overline{\zeta}_1^j - \overline{\xi}_1^i)( \overline{\zeta}_2^j - \overline{\xi}_2^i) (P_{t_0}^t)_{ij})\\
	\sum^{N}_{j=1} (( \overline{\zeta}_1^j - \overline{\xi}_1^i)( \overline{\zeta}_2^j - \overline{\xi}_2^i) (P_{t_0}^t)_{ij})& \sum^{N}_{j=1} (( \overline{\zeta}_2^j - \overline{\xi}_2^i)^2 (P_{t_0}^t)_{ij}).
\end{array}\right)
\end{eqnarray}
where  $[\overline{\zeta}_1^j, \overline{\zeta}_2^j]$ are the centers (mean  values) of each of the boxes, $B_j$ and $[ \overline{\xi}_1^i,  \overline{\xi}_2^i] = \sum^{N}_{j=1} ([\overline{\zeta}_1^j, \overline{\zeta}_2^j](P_{t_0}^{t_0 + \Delta t})_{ij})$. 

 \begin{figure} [!h]
\begin{center}
\includegraphics[height = 0.3 \textwidth, width = 0.7\textwidth]{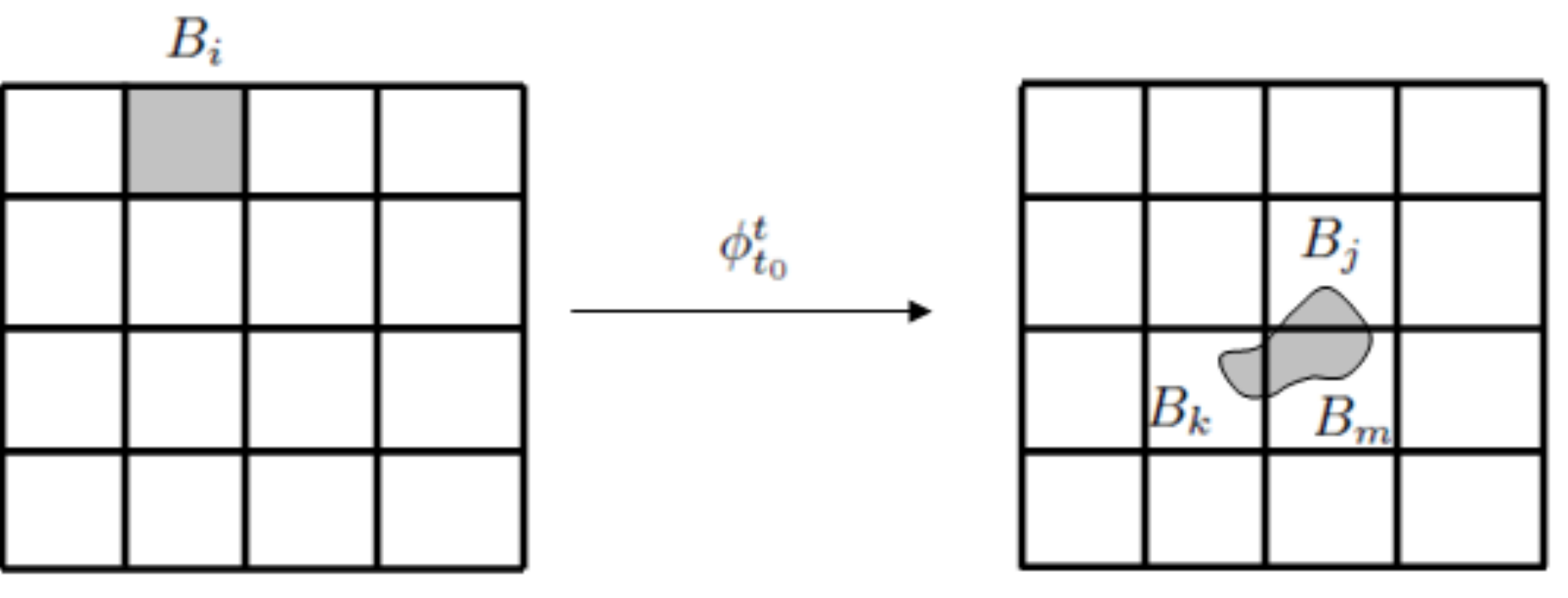}
\end{center}
\caption{\label{fig:approx_fig1}{\footnotesize{Illustration of the box discretization method. Box${B_i}$ with measure is mapped into three  boxes ${B_j}$, ${B_k}$ and ${B_m}$ by the flow map $\phi(B)_{t_0}^t$. The values of $(P_{t_0}^t)_{ij}$, $(P_{t_0}^t)_{ik}$ and $(P_{t_0}^t)_{im}$ are given by equations \ref{eq:P_mat}, \ref{eq:P_reverse} and \ref{eq:P_bar}}}}
\end{figure}

 \begin{figure} [!h]
\begin{center}
\includegraphics[height = 0.38 \textwidth, width = 0.7\textwidth]{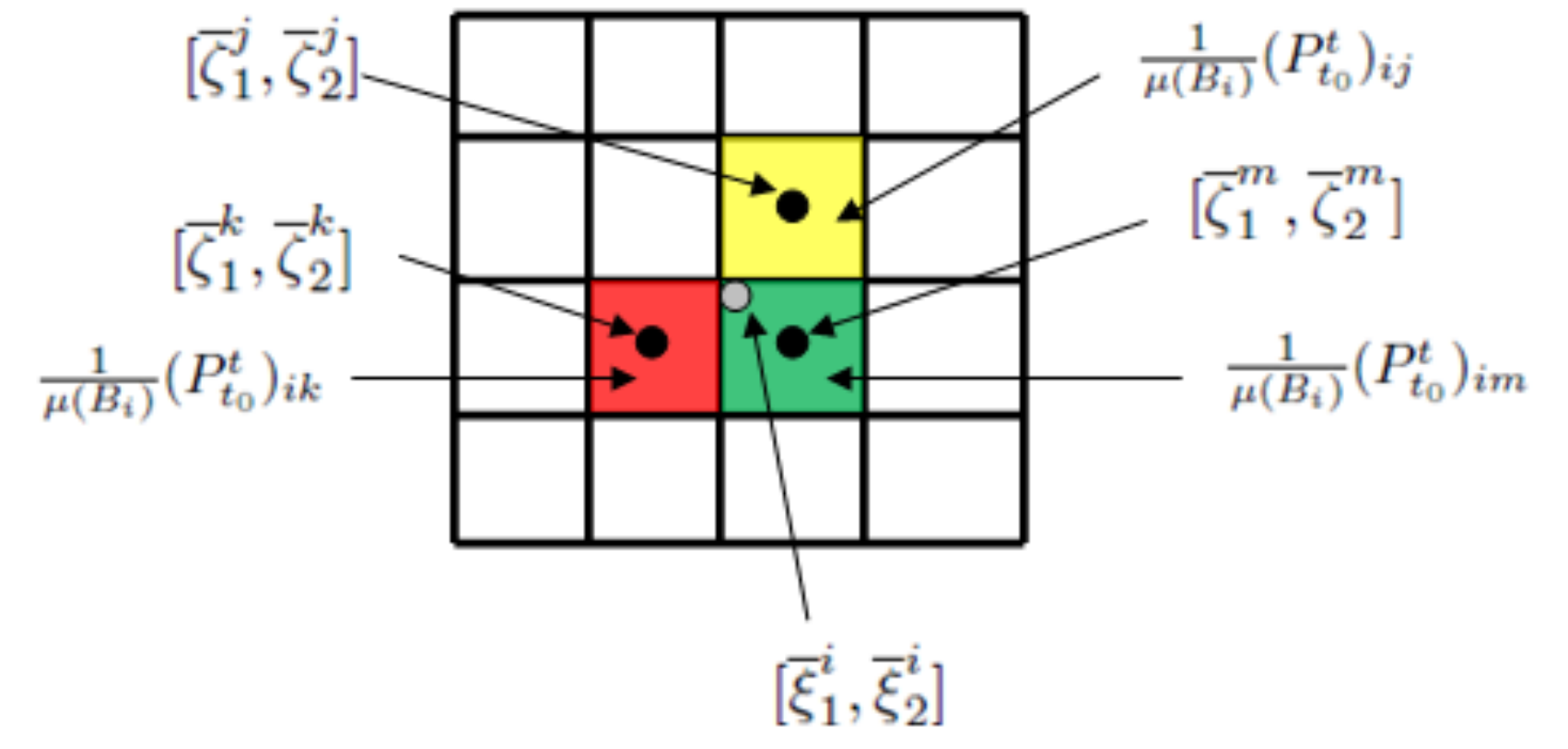} 
\end{center}
\caption{\label{fig:approx_fig2}{\footnotesize{Illustration of the calculation of the covariance, $I(v^i(t))$ in equation \ref{eq:covar_P_discrete}.}}}
\end{figure}
 
Intuitively the uniformly distributed points in box $B_i$ are mapped into some of the $N$ boxes by the flow map $\phi_{t_0}^t$, say boxes $B_j$, $B_k$ and $B_m$ as shown in figure \ref{fig:approx_fig1}. If the number of points that are mapped into each of the three boxes are $N_j$, $N_k$ and $N_m$, then each of the three sets $B_j \cap \phi_{t_0}^t(B_i)$, $B_k \cap \phi_{t_0}^t(B_i)$ and $B_m \cap \phi_{t_0}^t(B_i)$ are approximated by $N_j$, $N_k$ and $N_m$ uniformly distributed points in the boxes $B_j$, $B_k$ and $B_m$. The mean or average of these dispersed points is $[\overline{\xi}_1^i,  \overline{\xi}_2^i]$. The vector $v^i(t) =  \frac{1}{\mu(B_i)}[0, . . .,(P_{t_0}^t)_{ij}, 0, ..., (P_{t_0}^t)_{ik}, 0,  ..., (P_{t_0}^t)_{im}, ...0]$ with the nonzero values in columns $j$, $k$ and $m$. The mean or center of this set is given by $[\overline{\zeta}_1^j, \overline{\zeta}_2^j]$.  Proceeding thus, the covariance FTLE for each of the boxes in the domain can be found from equation \eqref{eq:covar_P_discrete}.

It should be emphasized that the computation of the FTLE using the Perron-Frobenius operator requires a set oriented definition of the FTLE. One cannot use the operator $P$  to track the evolution of sets of zero measure, since the integral on the right hand side of equation \ref{eq:Perron-Frobenius_op} is zero for this case. The standard method of FTLE which requires the evolution of individual trajectories passing through the nodal points of a mesh at the initial instant of time, does not make any direct reference to sets of positive  measure around these nodal points. Hence a set oriented definition of FTLE is necessary to take advantage of the Perron-Frobenius operator. In practice the computational approach for this new definition too has to make use of discrete initial conditions and trajectories , but sets of positive measure are explicitly modeled by these discrete initial conditions.
\section{Examples}
In this section we illustrate the computation of the FTLE using the Perron-Frobenius operator for two flows. The first is the lid-driven cavity flow studied for its mixing properties in \cite{Grover_Paper} and the double gyre flow that has been a prototype flow in the LCS and almost invariant sets literature, \cite{Shadden1}, \cite{Ross_Lekien}, \cite{Froyland2}.
\subsection{Lid driven cavity flow}
The problem of transport in the model of the lid driven cavity flow has been investigated in \cite{Grover_Paper} and is considered  here as the first example because of its simple piecewise steady velocity field. The flow is described by the stream function
\begin{equation}\label{eq:lid_flow}
\psi(x,y) = \displaystyle\sum\limits_{n=1}^2  U_n C_n f_n(y) \sin\left(\frac{n \pi x}{a}\right)
\end{equation}
defined on the domain $[0,a]\times[-b,b]$ for time $0\geq t \geq \tau_f/2$. where 
\begin{equation*}
f_n = \frac{2\pi y}{a} \cosh{\left(\frac{n\pi b}{a}\right)}\sinh{\left(\frac{n\pi y}{a}\right)} - \frac{2\pi b}{a} \sinh{\left(\frac{n\pi b}{a}\right)} \cosh{\left(\frac{n \pi y}{a}\right)}.
\end{equation*} 
and 
\begin{equation*}
C_n = \frac{a^2}{2n\pi^2b} \left[\frac{a}{2n\pi b}\sinh{\left(\frac{2n\pi b}{a}\right)} + 1\right]^{-1}
\end{equation*} 
For time $\tau_f/2 \geq t \geq \tau_f$, the sign of the velocity term $U_1$ is changed. This reflects the streamlines about $x=a$ after a time $\tau_f/2$.

Using symmetry arguments given in \cite{JieChen1} and \cite{JieChen2}, a specific ratio of the magnitudes of the terms $U_2/U_1$ along with a fixed value of the period of the flow $\tau_f$ is found such that it generates three period-3 fixed points in the domain $[0,a]\times[-b,b]$. The specific values of the constants that we borrowed from \cite{JieChen1} and \cite{JieChen2} are $U_1 = -1$ and $U_2 = 0.841298$ and $\tau_f^* /2 \approx 4.740202$ for the domain $[0,a] \times[-b,b] = [0,6] \times [-1,1]$.

A perturbation of the time period of the flow from the critical value of $\tau_f^* /2\approx 4.740202$ destroys the fixed points. The method of almost invariant sets was used in \cite{Grover_thesis} and \cite{Grover_Paper} to study mixing for different values of the perturbed time period. Since our main interest here is to illustrate the FTLE computed from the Perron-Frobenius operator, we choose a single case, a specific value of half time period $\tau_f/2 = 4.848$ for our study.

\begin{figure} [!h]
\begin{center}
\begin{tabular}{cc}
\includegraphics[height = 0.22 \textwidth, width = 0.51\textwidth]{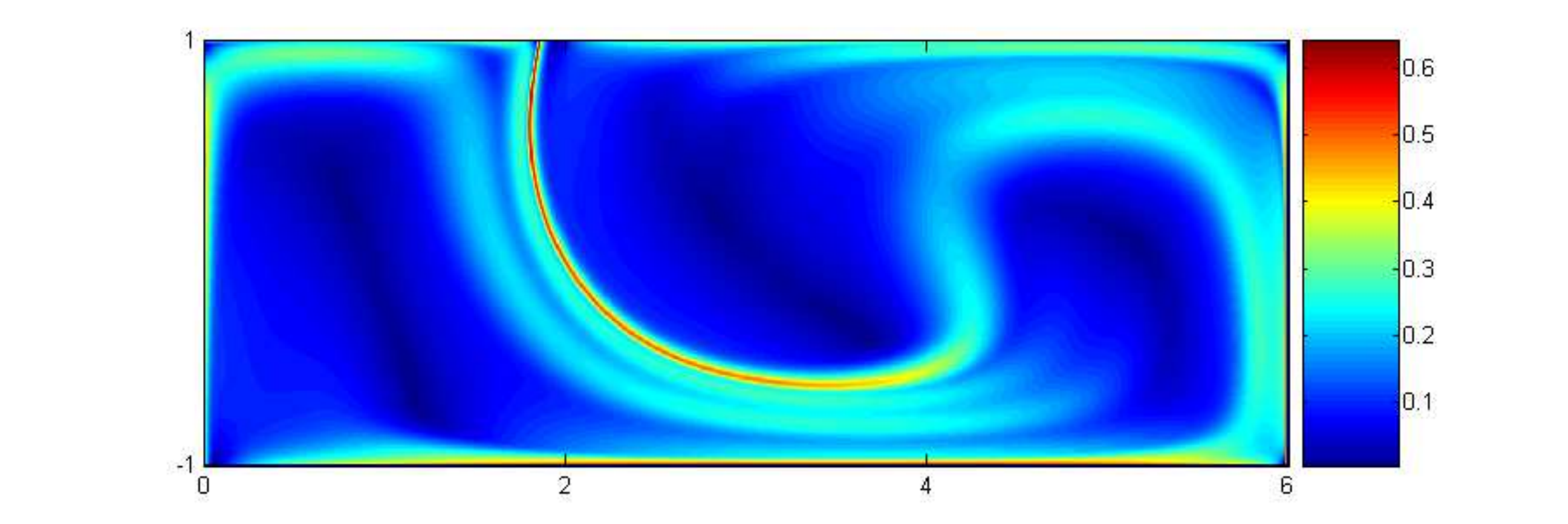} & 
\includegraphics[height = 0.22 \textwidth, width = 0.51\textwidth]{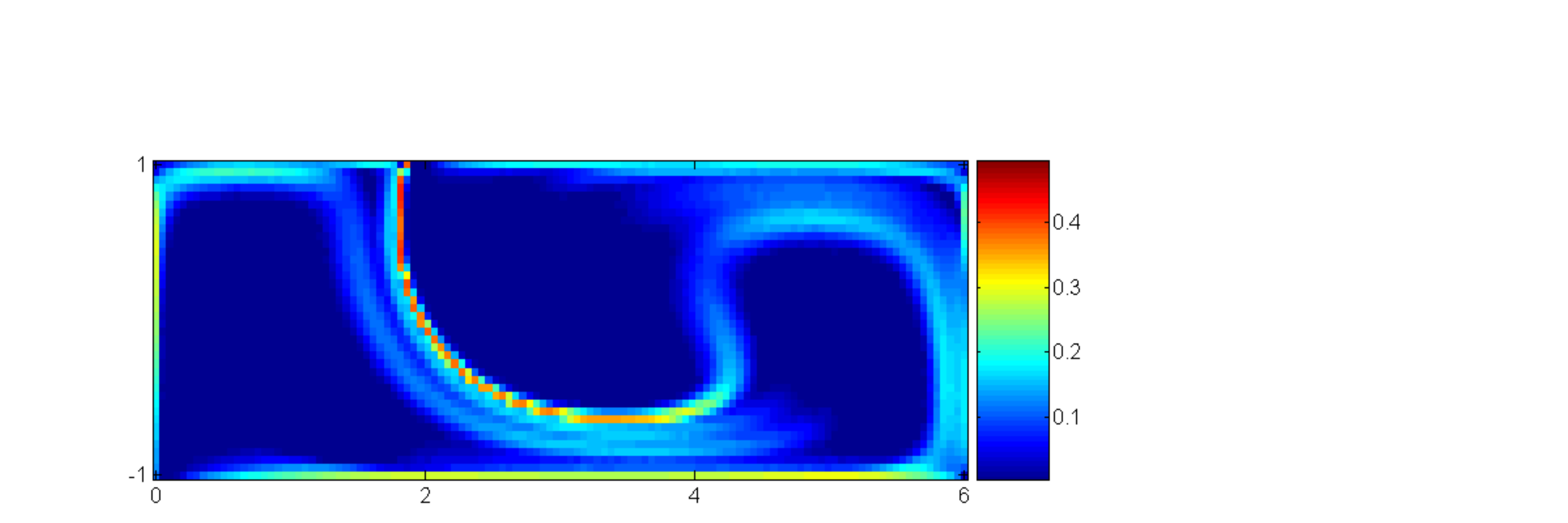} \\
\footnotesize {(a) $T = \tau_f$} & \footnotesize {(d) $T = \tau_f$}\\
\includegraphics[height = 0.22\textwidth, width = 0.51\textwidth]{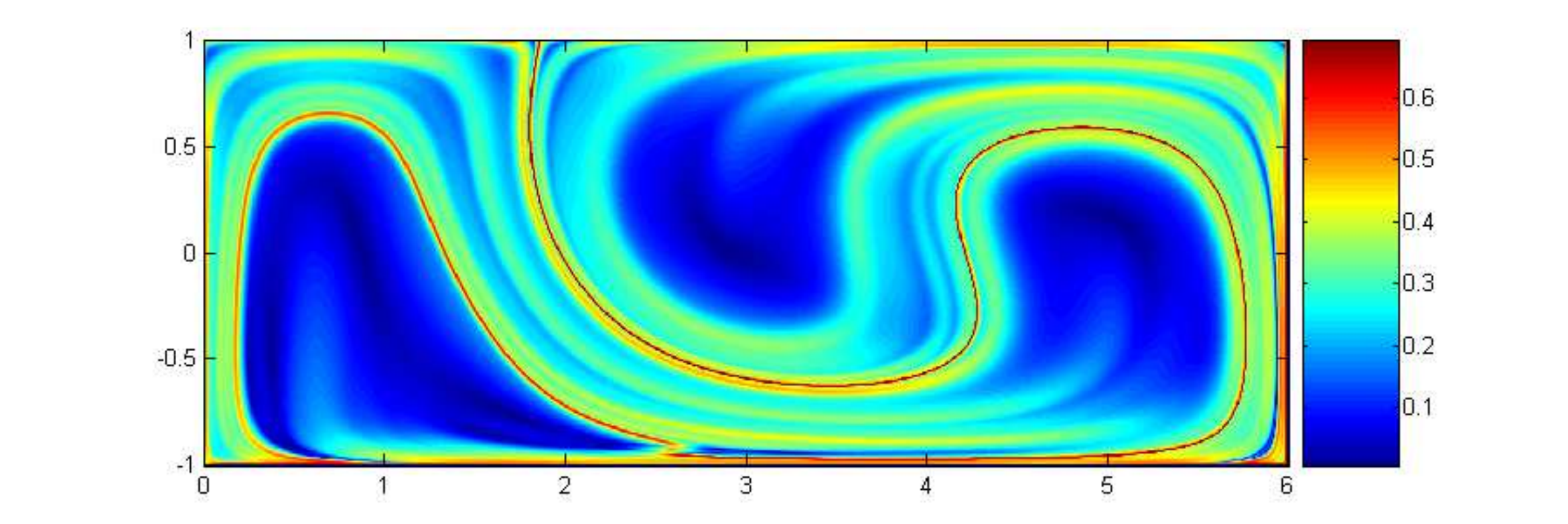}&
\includegraphics[height = 0.22 \textwidth, width = 0.51\textwidth]{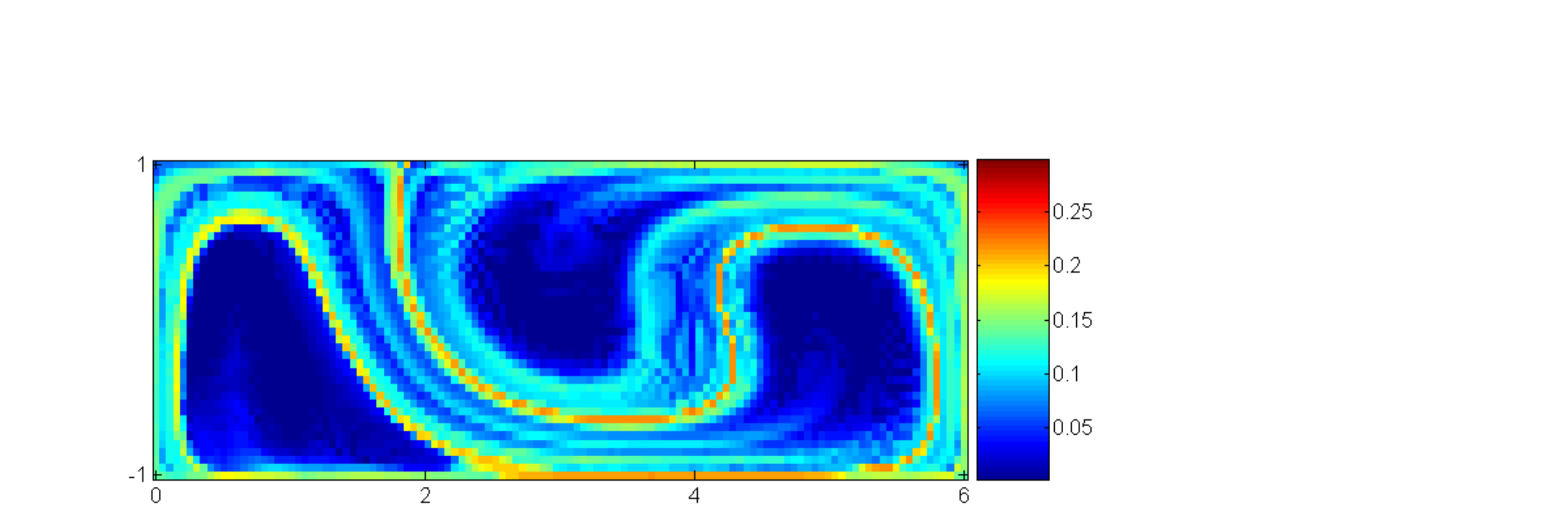} \\
\footnotesize {(b) $T = 2\tau_f$} & \footnotesize {(e) $T = 2\tau_f$}\\
\includegraphics[height = 0.22 \textwidth, width = 0.51\textwidth]{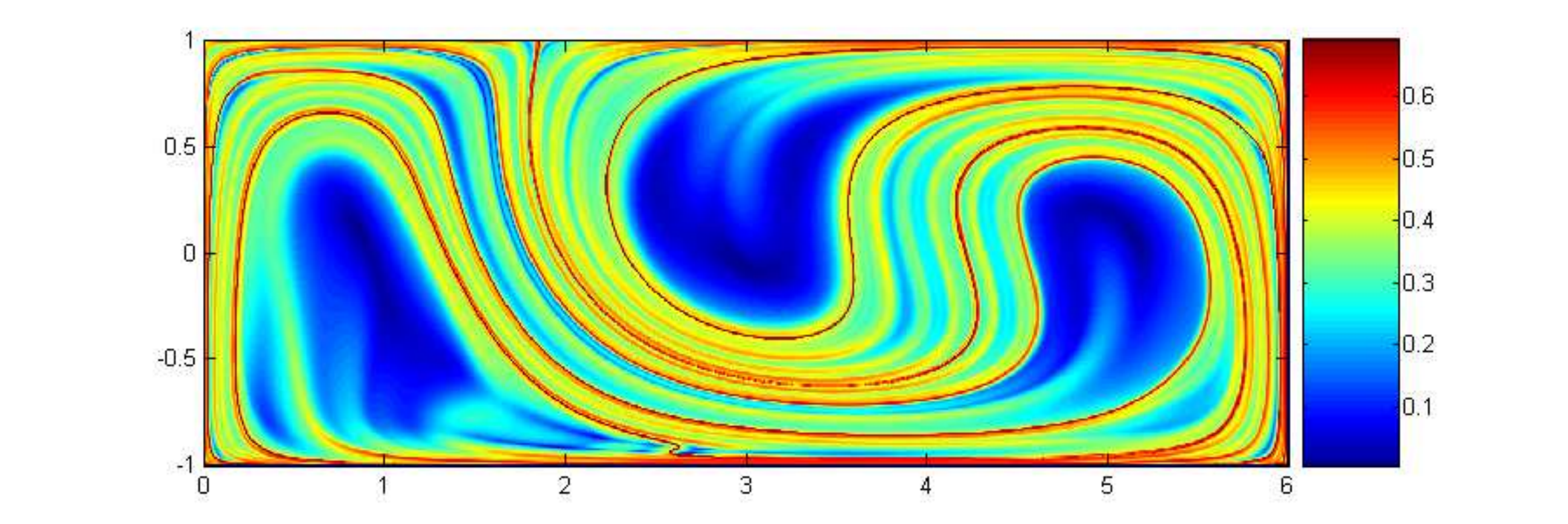} & 
\includegraphics[height = 0.22 \textwidth, width = 0.51\textwidth]{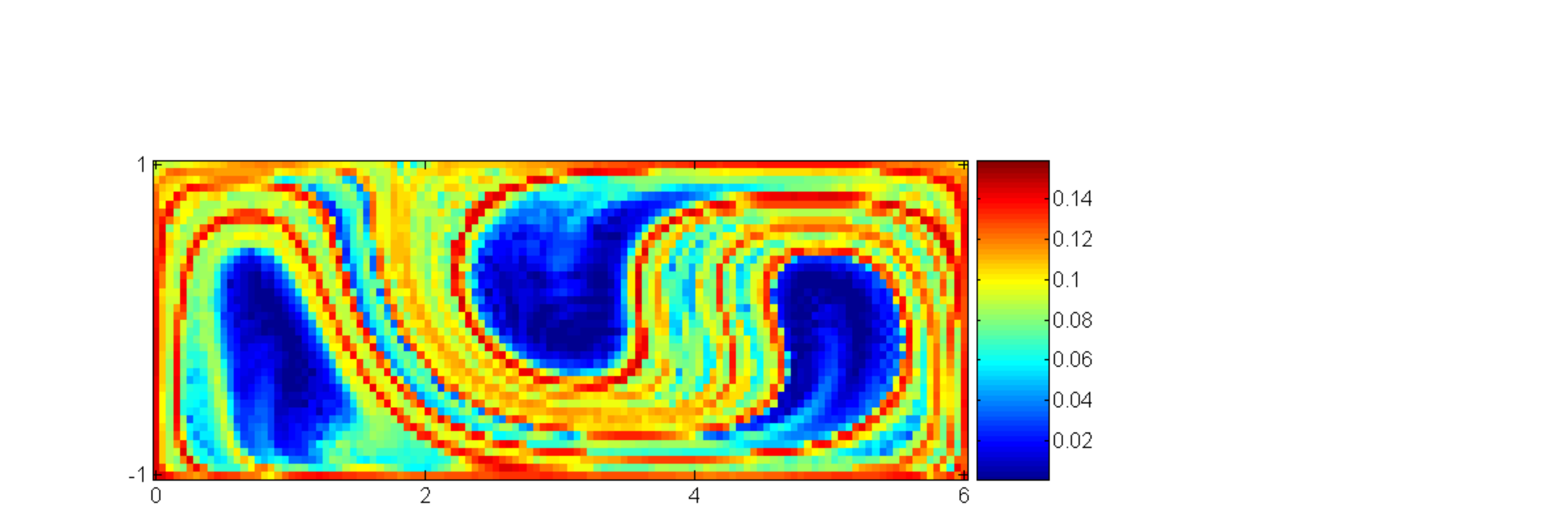} \\
\footnotesize {(c) $T = 3\tau_f$} & \footnotesize {(f) $T = 3\tau_f$}
\end{tabular}
\end{center}
\caption{\label{fig:ftle_lidflow}{\footnotesize{FTLE for integration time $T$ for the lid driven cavity flow. On the left (a-c) is  the FTLE field calculated using the standard method of stretching of line elements. On the  right (d-f) is the FTLE calculated using the operator $P$ and equation \eqref{eq:covar_P_discrete}.}}}
\end{figure}

The standard FTLE computation was done for the system for different integration times; $\tau_f$, $2\tau_f$ and $3\tau_f$, the results of which are shown in figure \ref{fig:ftle_lidflow}(a)-(c) by integrating points initially spaced at a distance of $0.05$.  The domain is then divided into 4800 boxes each box containing a 100 points and the time reversible Perron-Frobenius matrix $P_r$ is found by integrating a total of 480,000 initial points for a period equal to $\tau_f$. The vector space has $120 \times 40$ dimensions which are the discretized finite approximation of the infinite dimensional space of Lebesgue integrable functions. Uniform density functions each supported on a box are the bases vectors for the finite dimensional vector space. The covariance based FTLE computation was performed using the Perron-Frobenius operator $P_0^{\tau_f}$. Figure \ref{fig:ftle_lidflow} (d) is obtained by computing the covariance of the bases functions from the rows $P_0^{\tau_f}$, while figure \ref{fig:ftle_lidflow} (e) and (f) are obtained by computing the covariance  of the basis functions from the  rows of $P_0^{2\tau_f} = (P_0^{\tau_f})^2$ and $P_0^{3\tau_f} = (P_0^{\tau_f})^3$ respectively. 
 
\begin{figure} 
\begin{center}
\begin{tabular}{cc}
\includegraphics[height = 0.22 \textwidth, width = 0.51\textwidth]{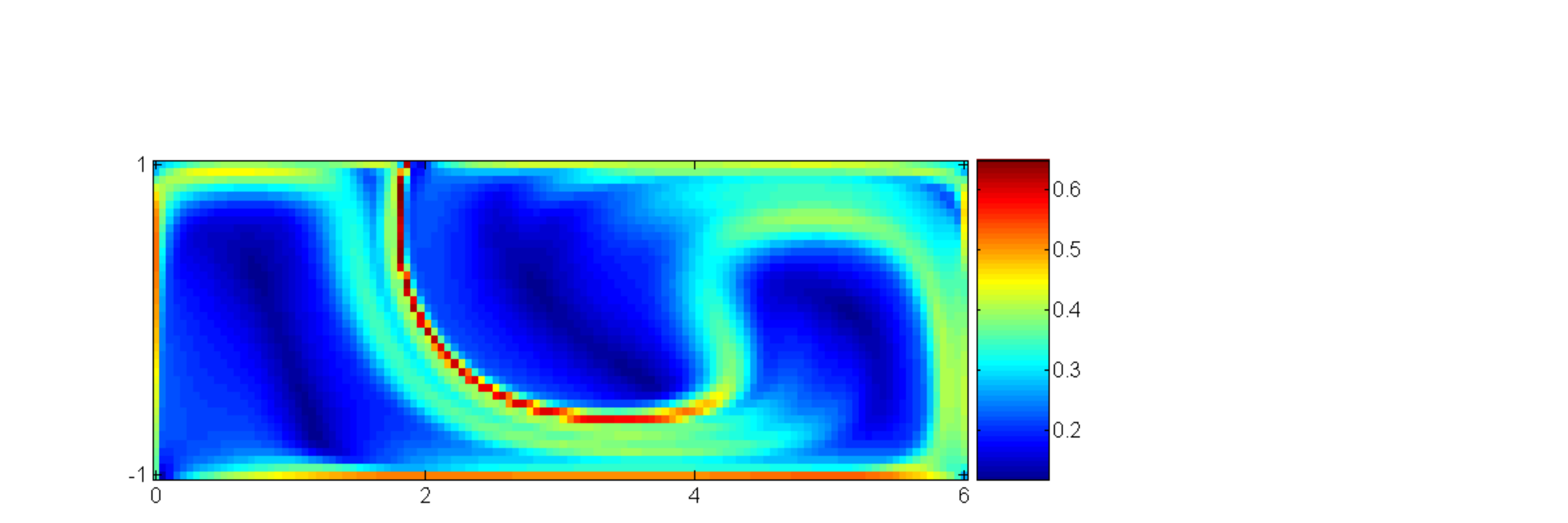} &
\includegraphics[height = 0.22\textwidth, width = 0.51\textwidth]{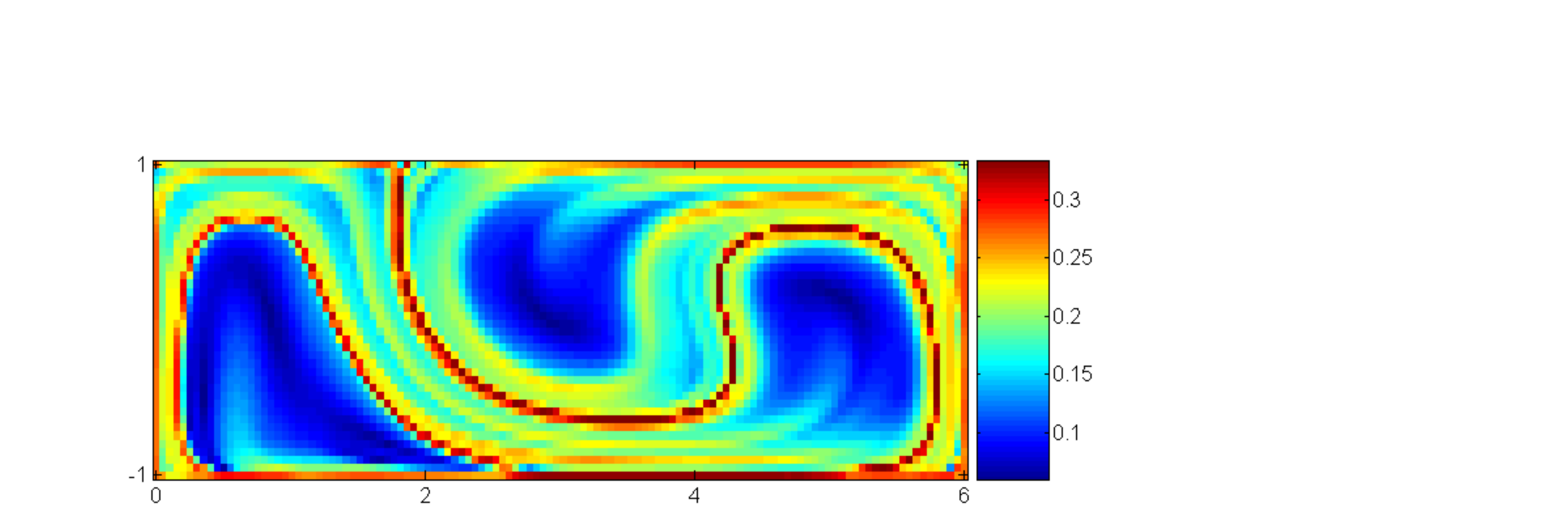}\\
\footnotesize {(a) $T = \tau_f$} & \footnotesize {(b) $T = 2\tau_f$} \\
\end{tabular}
\includegraphics[height = 0.22 \textwidth, width = 0.51\textwidth]{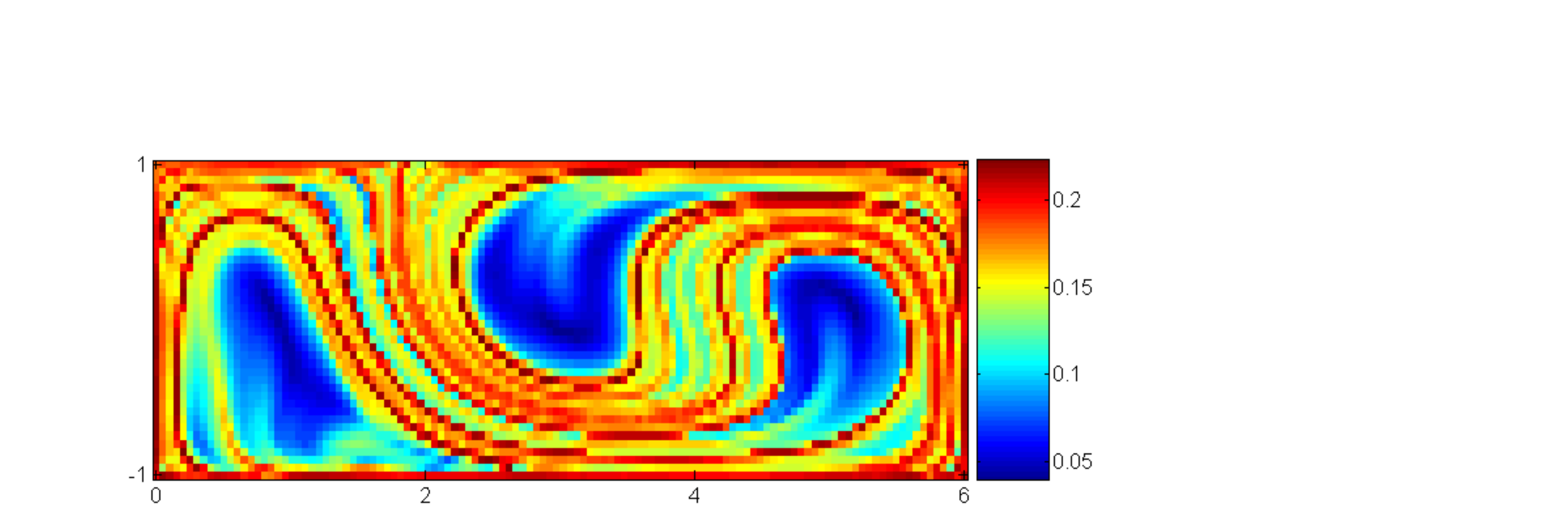} \\
\footnotesize {(c) $T = 3\tau_f$} \\
\end{center}
\caption{\label{fig:direct_ftle_lidflow}{\footnotesize{Covariance FTLE for lid driven cavity flow computed  by direct integration of initial conditions using equation \eqref{eq:covar_Pf} for various integration times.}}}
\end{figure}

It is evident from the figure \ref{fig:ftle_lidflow} that the FTLE field has the same ridge structure when computed by the line stretching approach or the covariance approach using the Perron-Frobenius operator. Computing the covariance based FTLE field directly by integrating all the initial conditions to $3\tau$, while more accurate is computationally more intensive. In fact the time  taken for this method increases  almost linearly with the integration time. The FTLE field computed from the Perron-Frobenius operator has the same features as the covariance FTLE field by integrating  the points in  each of the boxes for $3\tau$ and computing the covariance of the bases functions which is shown in figure \ref{fig:direct_ftle_lidflow}. 

While the  FTLE field computed directly by integration and that computed by the Perron-Frobenius  operator have the same ridge features, they differ in the magnitude of the FTLE field. This discrepancy in the magnitude is due to two reasons. The Ulam method of approximating the Perron-Frobenius operator by a matrix introduces approximations.The measure of each of the boxes is approximated by a finite number of discrete points introducing further errors. Lastly the calculation of the covariance matrix $I(v^i(t))$ using equation as explained in figure \ref{fig:approx_fig1} introduces  another level of approximation. As the box size in the computation is reduced and/or the number of points  per box is increased, it is natural to expect that the matrix approximation of $P$ and the covariance computation become more accurate, though no rigorous proof is offered here.

\subsection{Double gyre flow}
\begin{figure} [!h]
\begin{center}
\begin{tabular}{cc}
\includegraphics[height = 0.25 \textwidth, width = 0.51\textwidth]{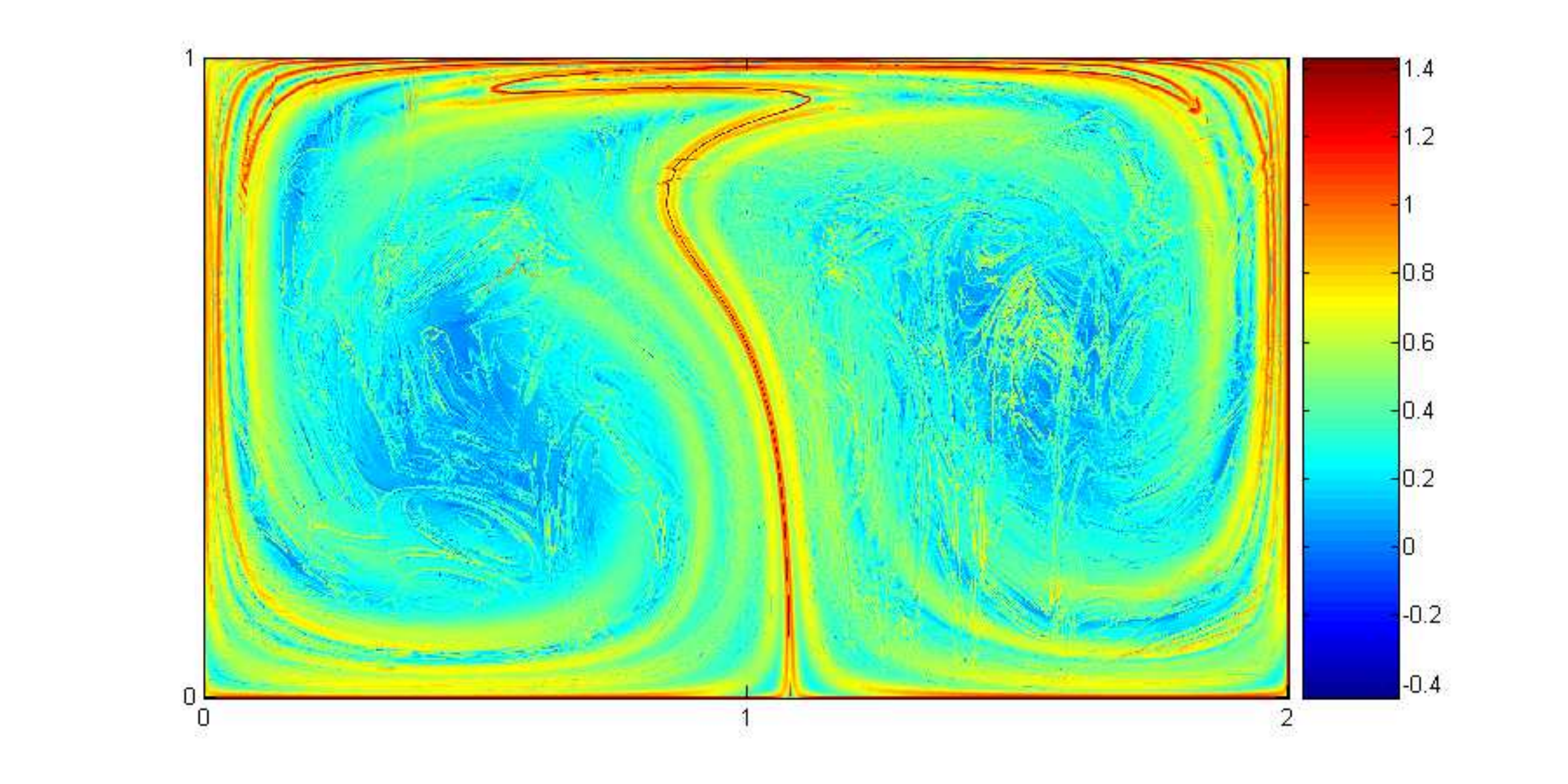} & 
\includegraphics[height = 0.25 \textwidth, width = 0.51\textwidth]{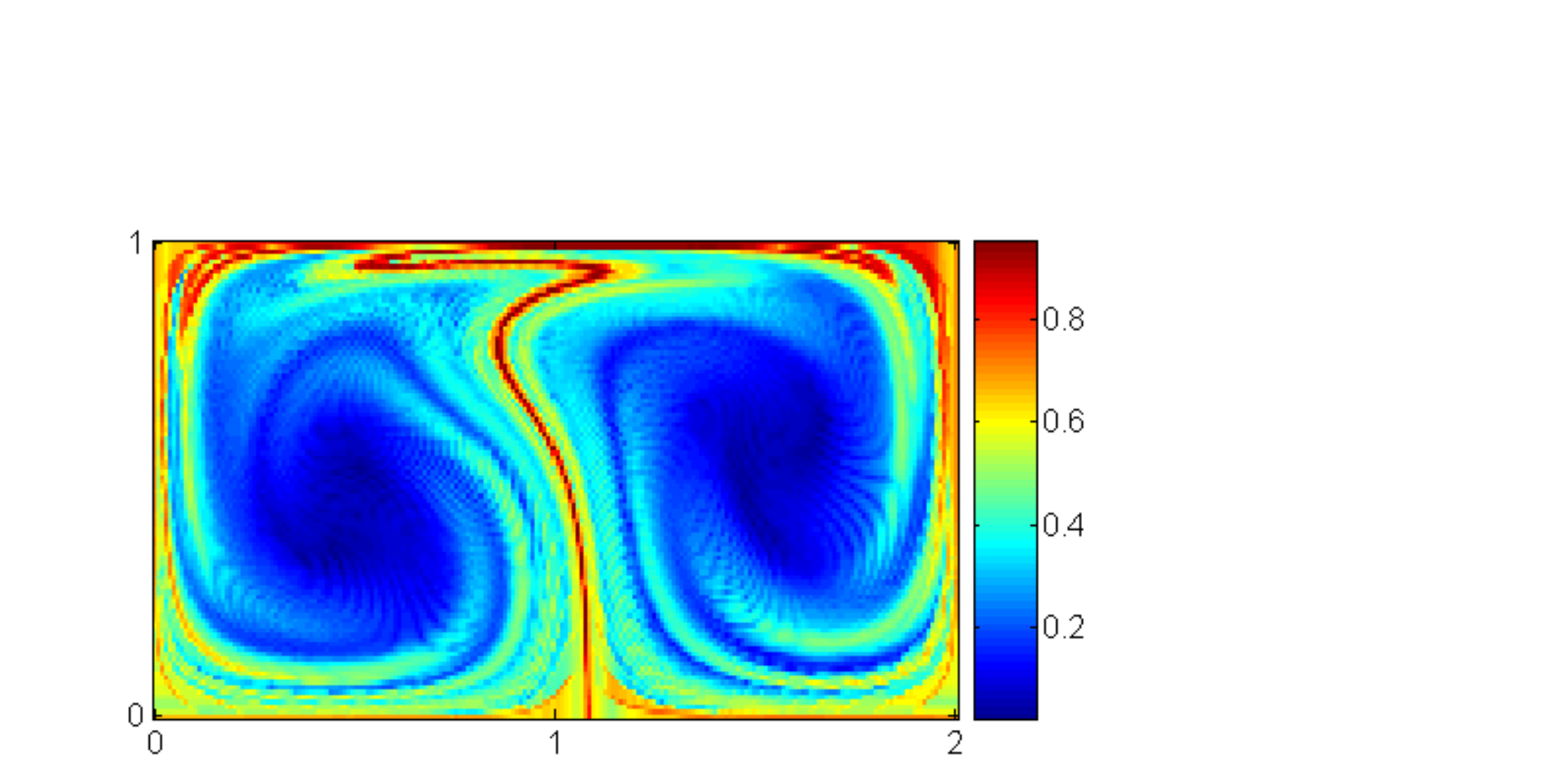} \\
\footnotesize {(a) $T = 5$} & \footnotesize {(b) $T = 5$}\\
\includegraphics[height = 0.25\textwidth, width = 0.51\textwidth]{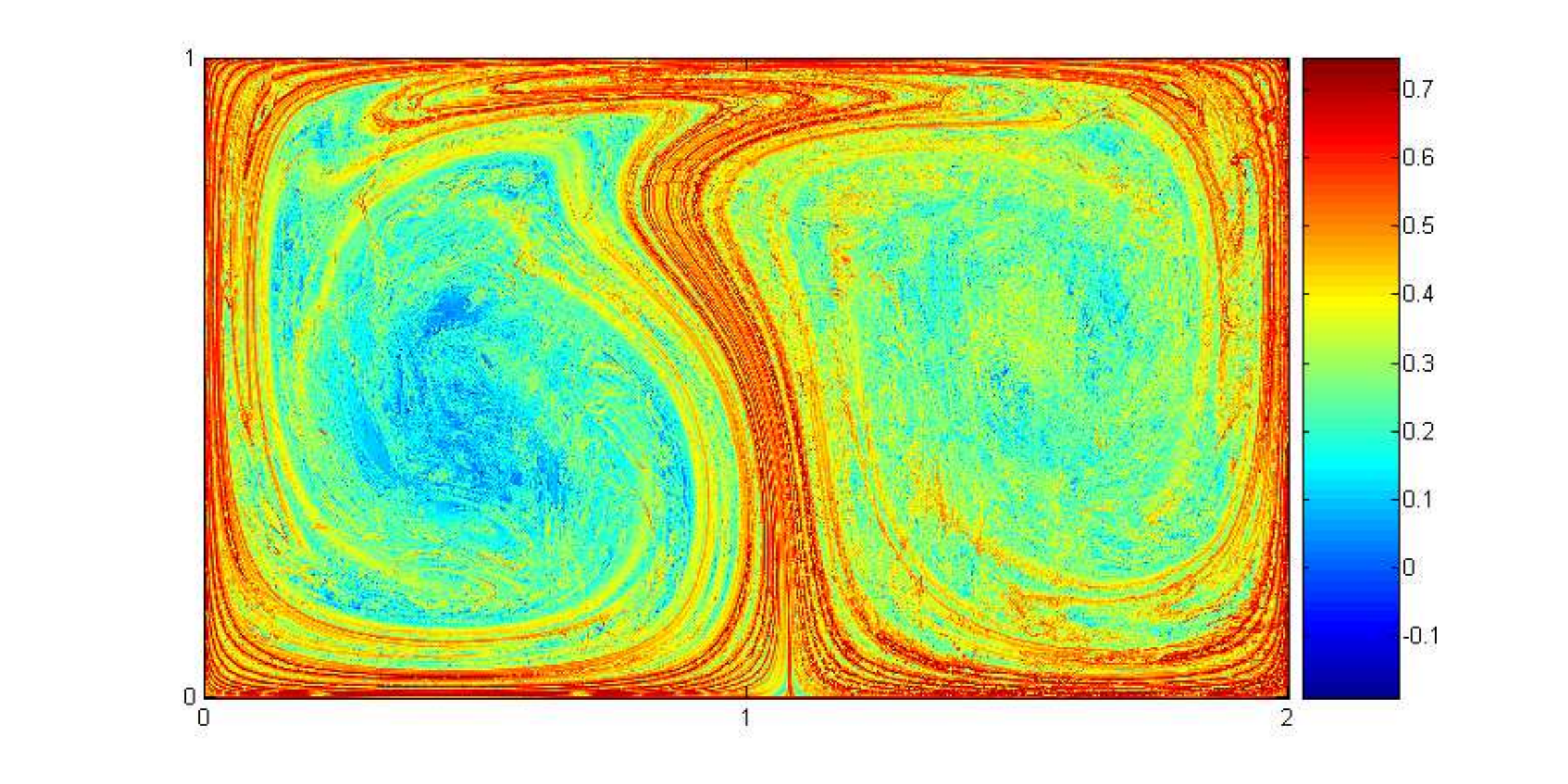}&
\includegraphics[height = 0.25 \textwidth, width = 0.51\textwidth]{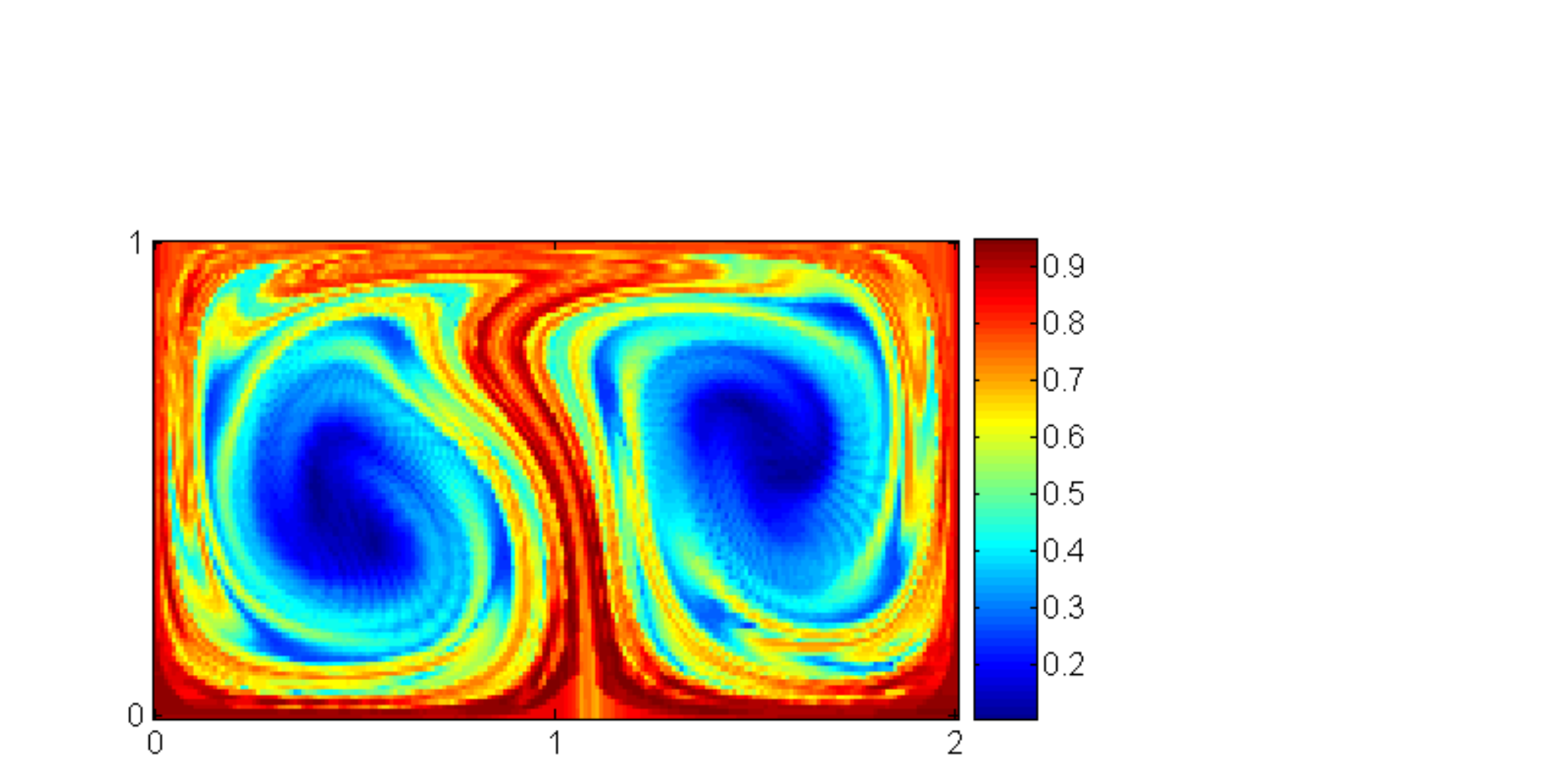} \\
\footnotesize {(c) $T = 10$} & \footnotesize {(d) $T = 10$}
\end{tabular}
\end{center}
\caption{\label{fig:ftle_doublegyre}{\footnotesize{FTLE for integration time $T$ for the double gyre flow. On the  left; (a) and (c) is the FTLE calculated using the standard method of stretching of line elements. On the  right; (b) and (d) is the FTLE calculated using the operator $P$ and equation \eqref{eq:covar_P_discrete}}}}
\end{figure}

The double gyre flow which is time dependent but periodic, has been one of the prototype flows in the LCS literature, \cite{Shadden1}, \cite{Froyland2}. The double gyre flow is defined by the stream function $\psi(x,y,t) = A \sin(\pi f(x,t))\sin{(\pi y)}$, with the parameters $A=0.25$, $\omega = 0.2\pi$ and $\epsilon = 0.25$.  The time period of the flow is $\tau = 1$.  The domain is discretized into 50000 square boxes of size $0.005$ each containing 625 points. The FTLE field computed using the stretching of line elements and the FTLE field computed from the Perron-Frobenius operator is shown in figure \ref{fig:ftle_doublegyre}. 

 \begin{figure} [!h]
\begin{center}
\includegraphics[height = 0.25\textwidth, width = 0.51\textwidth]{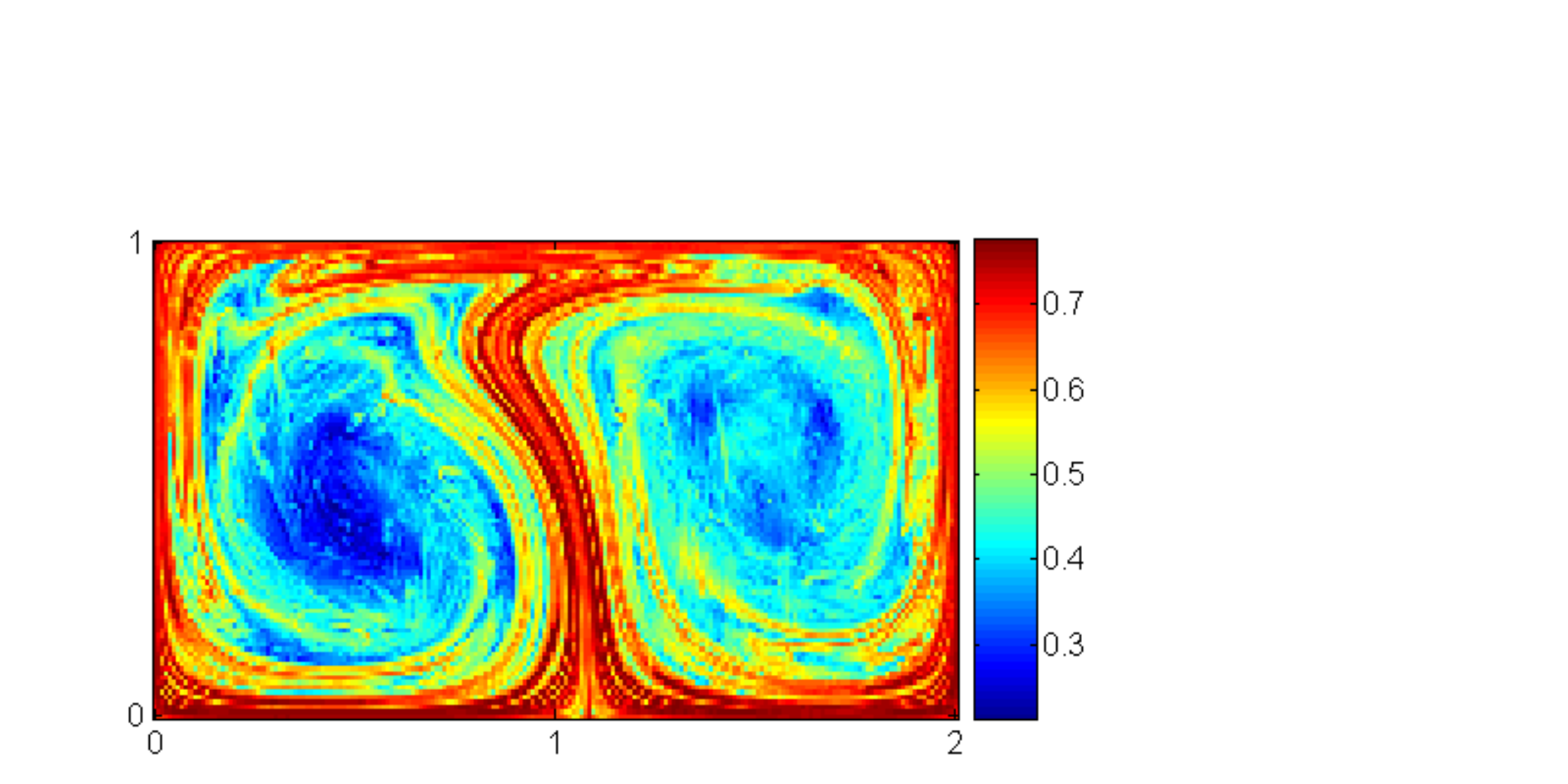}
\end{center}
\caption{\label{fig:ftle_direct_doublegyre}{\footnotesize{Covariance FTLE obtained by direct integration for an integration time $T = 10$ using equation \eqref{eq:covar_Pf} for the double gyre flow.}}}
\end{figure}

As observed in the discussion on the lid driven cavity flow, the FTLE field has the same ridge  features, when the covariance of the  sets $Pf$ are computed directly by integrating all the points in each box to $T=10$, as shown in figure \ref{fig:ftle_direct_doublegyre} and when computed using the Perron-Frobenius operator. Moreover because of the more accurate approximation of the domain by a higher number of boxes, the covariance FTLE field computed by a direct integration and that computed by the Perron-Frobenius operator are closer to each other in magnitude.

\section{Conclusion and Discussion}
A method to compute the FTLE field using the  Perron Frobenius operator has been introduced in this paper. To do this a modified definition of the FTLE was used which identifies deformation of sets of positive measure instead of the stretching of line elements. This methodology is general enough to be used in time-dependent flows and is particularly useful in the case of time-independent and periodic flows, where it offers a significant computational advantage by eliminating long time integrations.  The method of computing the covariance FTLE from Perron-Frobenius operator using equation \eqref{eq:covar_P_discrete} was illustrated by two examples; the lid  driven cavity flow and the double gyre flow. In both cases it was shown that the FTLE field computed by the new method has the same topological features as the FTLE field computed using the standard approach of line  stretching and  the direct computation of the covariance FTLE from the dispersion of points as in equation \eqref{eq:covar_Pf}. The computational time required for the calculation of the covariance FTLE  directly from the dispersion of points scales linearly with time in the most ideal  case. So the time taken to calculate $\sigma_I(t_0, t_0 + n\Delta t)$, for an integration time $n\Delta t$, is $n$ times  that of the time taken for calculating $\sigma_I(t_0, t_0 + \Delta t)$. Using the PF operator approach the time necessary to calculate the FTLE is almost independent of the multiple $n$. Thus the method proposed in this paper  is $n$ times computationally efficient.

Comparing  the method proposed in the paper, with that of the standard FTLE, the  computation of  the standard FTLE is faster. This  is because the standard FTLE computations were done a very coarse  grid in both the examples presented. Even though the time of integration is shortes, fewer initial conditions have to be integrated and thus the method is faster. However as has been pointed in the paper in section 3.1 this is achieved at the cost of erroneously ignoring the higher order terms in equation \eqref{eq:taylor_perturb}. The covariance based method of computing the FTLE does not use the linearized equations of a flow making it more general in scope of application. 

Moreover the covariance FTLE gives the deformation of sets a probabilistic interpretation. The new  method of computing the covariance FTLE introduced in this paper is based on the  Perron-Frobenius operator; a common tool in probabilistic methods of phase space transport. This puts the Perron-Frobenius operator at the center of study of phase space transport. This is perhaps intuitively obvious since the Perron-Frobenius operator contains all the information of global transport of sets.  The method of computing the covariance FTLE using the Perron-Frobenius operator is a step towards combining the probabilistic and geometric methods of  phase space transport, into a common unified framework.
%

\end{document}